\documentclass[12pt]{article}

\usepackage{amsmath}
\usepackage{amsfonts,amssymb,amsthm,overpic}
\makeatletter
\newcommand{\diam}{\mathop{\operator@font diam}}
\usepackage{setspace}
\usepackage{multicol}
\usepackage{multirow}
\usepackage[compact]{titlesec}

\usepackage{epsfig}

\newtheorem{theorem}{Theorem}[section]

\begin{document}

\title{\Huge{\textsc{On the Possibility of Singularities on the Ambient Boundary}}}

\author{Kyriakos Papadopoulos\\
\small{Department of Mathematics, Kuwait University, Kuwait}}

\date{}

\maketitle

\begin{abstract}
The order horismos induces the Zeeman $Z$ topology, which is coarser
than the Fine Zeeman Topology $F$. The causal curves in a spacetime
under $Z$ are piecewise null. $F$ is considered to be the most physical
topology in a spacetime manifold $M$, as the group of homeomorphisms
of $M$ is isomorphic to the group of homothetic transformations of $M$.
$Z$ was used in the Ambient Boundary-Ambient Space cosmological model,
in order to show that there is no possibility of formation of spacetime
singularities. In this article we question this result, by reviewing
the corresponding articles, and we propose new questions towards the improvement
of this model.
\end{abstract}

\section{Introduction.}

As a result of his paper \cite{Zeeman2}, Zeeman (see \cite{Zeeman1}) asked whether there is a topology, on the Minskowski spacetime $M$,
with a physical meaning which is related to the light cones
and is also more natural than the ordinary one. He answered to this question by defining
the Fine Topology (in this article and in \cite{Ordr-Ambient-Boundary} we call it the Fine Zeeman
Topology and we denoted it by $F$). Under this topology, for $M$, the group of homothetic
symmetries coincides with the group of homeomorphisms of $M$. Zeeman then conjectured
that this result can be extended to any curved spacetime, something which was proved to be
correct by G{\"o}bel (see \cite{gobel}).

The basic idea of ambient cosmology (see \cite{Ambient-Cosmology-Spacetime-Singularities} and \cite{Topology-Ambient-Boundary-Convergence}) is to regard our $4$-dimensional spacetime
as a bounding hypersurface, the conformal infinity, of a new cosmological metric
in $5$-dimensional ambient space. This ambient metric should not be confused with the
construction in \cite{Fefferman}, as it imposes a completely different condition on
the ``ambient'' boundary.

For a detailed exposition of the Ambient Boundary-Ambient Space model
we refer to \cite{Ambient-Cosmology-Spacetime-Singularities}. This model
has been proposed towards a reformulation of the conditions of the (standard)
singularity theorems of general relativity. A topological study in this
model resulted to article \cite{Topology-Ambient-Boundary-Convergence},
where is our main focus in this current article.

\section{Nominees for Topologies on the Ambient Boundary.}

In \cite{Topology-Ambient-Boundary-Convergence} the authors talk about
the uniqueness of the Zeeman Fine Topology and denote it by $Z$. For
convenience, as we have already mentioned in article \cite{Ordr-Ambient-Boundary},
we will denote this topology by $F$.

G{\"o}bel proved in \cite{gobel} a conjecture by Zeeman (see \cite{Zeeman1})
that for any spacetime, the group of homothetic symmetries must be isomorphic to
group of all homeomorphisms of the topology $F$ and, in addition, $F$ is the
unique topology having this property (all other have homeomorphism groups
isomorphic to the conformal group).

However, the authors of \cite{Topology-Ambient-Boundary-Convergence} actually
consider a coarser Zeeman topology that here we call $Z$ (as it has been also discussed in article \cite{Ordr-Ambient-Boundary}). The authors state that
 to describe $Z$ (which is called, by mistake, as the Zeeman Fine Topology in \cite{Topology-Ambient-Boundary-Convergence})
 ``we say that for $x\in M$ an open ball has the form $B_Z(x;r)= (B_E(x;r)-N(x)) \cup \{x\}$,
 where $B_E(x;r)$ is the Euclidean-open ball and $N(x)$ the null cone at $x$''. This is, in fact,
 is coarser than the Fine Zeeman Topology $F$ but finer than the Euclidean Topology $E$,
 resembling the following properties: (i) it is not locally homogeneous and the
 light cone through any point can be deduced from this topology, (ii) the group
 of all homeomorphisms of this topology is generated by the inhomogeneous Lorentz
 group and dilatations and (iii) it induces the $3$-dimensional Euclidean topology
 on every space axis and the $1$-dimensional Euclidean topology on every time axis,
 (see \cite{Zeeman1}). The following theorem though does not hold
 under $Z$:
\begin{theorem}Let $f : I \to Z$ be a continuous map of the unit interval $I$
 into $Z$. If $f$ is strictly order-preserving, then the image $f(I)$ is a
 piecewise linear path, consisting of a finite number of intervals along time axes.
 \end{theorem}

 So, as Zeeman states, ``although $Z$ is technically simpler than $F$ in having
 a countable base of neighbourhoods for each point, it is intuitively less attractive
 than $F$''.

We remark though that there is no restriction in the ambient boundary-ambient
space model that avoids the use of $F$. The question is why should one use
$Z$, given Theorem 2.1 and the remark of Zeeman that $Z$ is intuitively less
attractive than $F$, while all mentioned properties (i)-(iii) hold for $F$
as well as Theorem 2.1.

 Thus, the conclusion that there is impossibility of singularities and so it
is impossible to formulate the singularity theorems and basic causality results
refer to the topology $Z$ and not to $F$. ``For a sequence of causal curves
to converge to a limit curve one uses in an essential way the Euclidean balls
with their Euclidean metric and their compactness to extract the necessary
limits'', the authors state, something which is valid with the topology
$F$, as it contains Euclidean balls as well.

Hence, for the bounding spacetime $M$, the conformal infinity of the ambient space
$V$, it appears that there is a freedom of choice of either the standard Euclidean
metric topology, giving the usual manifold topology, or the Fine Zeeman topology $F$.

\section{Discussion, Conclusions and Open Questions.}

In our opinion one should focus on the results of the article \cite{Ordr-Ambient-Boundary}.
The authors there answer to the orderability problem (see \cite{Good-Papadopoulos} and \cite{Orderability-Theorem})
in the special case of the order {\em horismos}, $\rightarrow$ (see \cite{Penrose-Kronheimer} and
also \cite{Penrose-difftopology}), and prove that its induced
topology is the coarser Zeeman topology $Z$. Thus, in a spacetime equipped with $Z$
the causal curves are piecewise null; ``although time continues to be one of the coordinates,
it loses any relation to the idea of causality as it is known in general relativity'' (see
the Discussion in \cite{Ordr-Ambient-Boundary}. These conclusions refer to spacetime manifolds
under the topology $Z$, but are not unique for the ambient boundary-ambient space, as
there is the choice of $F$ as well.

Our main conclusion in this article is that one should review the results of
the article \cite{Topology-Ambient-Boundary-Convergence} as well as the discussion
of article \cite{Ordr-Ambient-Boundary}. In the first case, one should question
again the construction of the ambient boundary-ambient space model, given that
through $F$ there can be formed sufficient conditions for singularity formation
using convergent families of causal curves. In the second case, one could question
the definition of dynamical evolution on a spacetime under $Z$ to a spacetime under $F$ (see \cite{Ordr-Ambient-Boundary}).

\end{document}